\documentclass[useAMS,usenatbib]{emulateapj} 

\usepackage{graphicx}
\usepackage{ulem}
\usepackage{amssymb}
\usepackage[%
    colorlinks = true,%
    linkcolor = blue,%
    urlcolor  = blue,%
    citecolor = webgreen,%
    anchorcolor = blue]{hyperref}

\newcommand{\ufhref}[3][blue]{\href{#2}{\color{#1}{#3}}}%

\usepackage[dvipsnames]{color}

\shorttitle{The mysterious  H$\alpha$ absorption line}
\shortauthors{Sethi, Shchekinov, Nath}

\begin{document}


\newcommand{\3}{\ss}
\newcommand{\n}{\noindent}
\newcommand{\eps}{\varepsilon}
\def\be{\begin{equation}}
\def\ee{\end{equation}}
\def\ba{\begin{eqnarray}}
\def\ea{\end{eqnarray}}
\def\de{\partial}
\def\msun{M_\odot}
\def\div{\nabla\cdot}
\def\grad{\nabla}
\def\rot{\nabla\times}
\def\ltsima{$\; \buildrel < \over \sim \;$}
\def\simlt{\lower.5ex\hbox{\ltsima}}
\def\gtsima{$\; \buildrel > \over \sim \;$}
\def\simgt{\lower.5ex\hbox{\gtsima}}
\def\etal{{et al.\ }}
\def\red{\textcolor{red}} 
\def\blue{\textcolor{blue}}

\definecolor{webgreen}{rgb}{0,.5,0}
\definecolor{webbrown}{rgb}{.6,0,0}


\title{The mysterious $6565\,\AA$ absorption feature of the Galactic halo} 

\author{Shiv K. Sethi$^1$, Yuri Shchekinov$^{1,2}$, Biman B. Nath$^1$}
\affil{$^1$Raman Research Institute, Sadashivanagar, Bangalore 560080, India\\
$^2$ Lebedev Physical Institute, 53 Leninsky Ave, Moscow, 119991, Russia}

\begin{abstract}
We consider various possible scenarios to explain
the recent observation of a claimed broad H$\alpha$ absorption in our Galactic halo with peak optical depth $\tau \simeq 0.01$ and equivalent width $W \simeq 0.17 \, \rm 
\AA$. We show that the absorbed feature cannot arise from the circumgalactic and ISM  H$\alpha$ absorption. As the observed absorption feature is quite broad 
$\Delta\lambda \simeq 30 \, \rm \AA$, we also consider  CNO lines that lie close to H$\alpha$ as   possible alternatives to explain the feature. We show such lines could 
also not account for the observed feature.  Instead, we suggest that it can arise  from diffuse interstellar bands (DIBs) carriers or polyaromatic hydrocarbons (PAHs) absorption. While we identify several such lines
close to the H$\alpha$ transition, we are unable  to determine  the molecule
responsible for the observed feature, partly  because of  selection 
effects that prevents us from identifying DIBs/PAHs features close to  H$\alpha$ using local observations. Deep integration  on a few extragalactic sources
with high spectral resolution might allow us to distinguish between different
possible explanations. 
\end{abstract}

\keywords{HI: absorption -- diffuse radiation -- intergalactic medium --
quasars: general}

\section{Introduction}
A recent observation of claimed  H$\alpha$ absorption along the lines of sight to SDSS galaxies  has put the physical
state of hydrogen  atoms in the Galactic halo in a spotlight.  \cite{zhang2017} [ZZ17 hereafter] stacked the spectra of more than
700,000 galaxies from the Sloan Digital Sky Survey.  Focussing on the wavelength range of $6340\hbox{--}6790\, \rm \AA$, they
detected a broad H$\alpha$ absorption line with peak optical depth $\tau \simeq 0.01$ and   equivalent width $\sim 0.17\, \rm \AA$.
The correlation between the features  and  Galactic longitude is consistent with the absorbing systems 
being located at rest in our Galactic halo within $L\sim 100$ kpc.

The measured width of the line  corresponds to a line of sight velocity dispersion of order $\pm 700 \, \rm  km \,  s^{-1}$, but
there is significant uncertainty in this interpretation, and the velocity spread can be of order $\pm 390 \, \rm  km \, s^{-1}$ (ZZ17). The absorption line is consistent with an  average column density of hydrogen atoms in $n=2$ state: 
$N_2\approx(7.34 \pm 0.04) \times 10^{11}$ cm$^{-2}$. The mean absorption map suggests that the absorption is isotropic 
and is prevalent across most lines of sights.

In the next section, we discuss  possible implications of interpreting 
the observed feature as H$\alpha$ absorption;  we discuss  direct constraints and possible mechanisms to populate the $2p$ and $2s$ states. In section~\ref{sec:dibpah}, we consider physical processes that might explain the observation while obviating the direct constraints. We consider trapped Ly-$\alpha$ lines, metal lines, and complex molecules (DIB carriers and PAH) as possible candidates.

\section{Maintaining H$\alpha$ absorbing circumgalactic gas} \label{sec:halpha}
The crux of the problem is to maintain a large amount of H$\alpha$ absorbing gas in the halo, which can
explain the observed value of column density $N_2$. There are 
two aspects to consider in order to understand this observation: pumping HI($n=2$) states to explain the observed column density, and constraining emission from the excited states. 

The observed absorption could be caused 
by either  $2p$ or $2s$ state. The H$\alpha$ transition could 
arise from multiple transitions:  
two transitions originating from the $2p$ level, $2p \hbox{--} 3d$ and $2p \hbox{--} 3s$,
or by a $2s\hbox{--}3p$ transition\footnote{https://physics.nist.gov/\\PhysRefData/Handbook/Tables/hydrogentable3.htm}. ZZ17
reported an absorption trough with peak optical depth $\tau \simeq 0.01$ and equivalent width $W = 0.170 \pm 0.001$\AA\,, yielding a column density  $N_2 = 1.13 
\times 10^{20} W/(f\lambda^2) \, \rm \AA \, cm^{-2}$ with an oscillator strength $f = 0.6410$.  In principle, we could  derive multiple column densities depending 
on oscillator strengths of these  transitions but that does not change the conclusions we reach in the paper. 

One can readily rule
out the  $2p$ level if the absorption originates in a medium optically thin
to Ly-$\alpha$ photons.  The corresponding
Ly-$\alpha$ photon energy density in the halo due to radiative de-excitation of $2p$ level would be $N_2 A_{2p,1s} (10.2 \, {\rm eV})/c \approx
10^{11} \, \rm eV \, cm^{-3}$, which is more than a factor $\simeq 10^{12}$
larger than the observed value  of UV background in the solar neighbourhood \citep[see, e.g. chapter 12 in ][]{2011piim.book.....D}. Therefore,
we consider the $2s$ state. The decay of $2s$ state gives two photons with
a broad energy distribution \citep{sptz}. As the A-coefficient of 
this decay is nearly 8 orders of magnitude smaller than Ly-$\alpha$ transition it causes  a corresponding decrease in the photon energy density. Even this decay gives a  photon number density which  larger than the observed value by
a factor $10^4$,  but we consider it as a plausible mechanism
to underline  the possible physical processes that can populate the first
excited state of hydrogen in the ISM. 

\subsection{2s state} 
 As the $2s$ state is metastable with a two-photon decay rate $A_{2s,1s}=8.227$ s$^{-1}$ \citep{sptz}, 
keeping the  excited state sufficiently populated is much easier in this case  as compared to  
the $2p$ state. There are several mechanisms of populating the  $2s$ state: radiative recombination of ionized hydrogen  to the $2s$ state, collisional 
excitations by thermal electrons  (contributions from protons and other ions is negligible)  and radiative excitations of higher energy levels $n\geq 3$ by  Lyman-series photons followed by radiative decays to $2s$ state.  

The equation of the balance of 2$s$ level is given by:
\be \label{pop2s}
n_{2s}={\alpha_2(T)\over A_{2s,1s}}n_e^2+{\Gamma_{1s,2s}+C_{1s,2s}(T) n_e\over A_{2s,1s}}n_{1s},
\ee
where $C_{1s,2s}(T)$,  $\alpha_2(T)$ is the coefficient of 
collisional excitation  \citep[e.g.][]{janev,flin,mclaugh} and    recombination  to the $2s$ state, respectively\footnote{ collisional excitation/recombination to states $n \ge 3$ followed by radiative decay to $2s$ state   have 
negligible contribution.}. $\Gamma_{1s,2s}$ is the rate at which Lyman series photons can populate the $2s$ state; $\Gamma_{1s,2s} = \sum_{i>2}\Gamma_{1s,i}P_{i2}$. Here
$\Gamma_{1s,i}$ gives the rate at which absorbed photons populate states with $i=n>2$
and $P_{i2}$ is the probability that through radiative decay the atom returns
to the $2s$ state, e.g. $P_{32} = A_{32}/(A_{31}+A_{32})$. The  absorption of photons by the $3p$ state dominates this process.

\subsubsection{Recombination} 
This process will dominate if the medium is highly ionized or $n_e \simeq  n_{\rm H}$, where $n_H$ is the total hydrogen density. Assuming a photoionized region with $T= 10^4 \, \rm K$, this gives:
\be \label{rec}
n_{2s}\simeq {\alpha_2(T) n_e^2\over A_{2s,1s}}\simeq  10^{-12} \left ({n_e\over 10 \, \rm cm^{-3}}\right)^2 \, {\rm cm^{-3}}, 
\ee 
To satisfy the observation, $n_{2s} \simeq N_2/L$, we require the size of the region $L\simeq 100 \, \rm kpc$ for $n_e \simeq 10 \, \rm cm^{-3}$. 
This  results in an   emission measure (EM): 
\be 
EM\simeq A_{2s,1s}N_2/\alpha_2(T)\simeq  10^7~{\rm ~cm^{-6}~pc}. 
\ee
For comparison the Orion  nebula has an emission measure $\simeq 5 \times 10^6 \, \rm cm^{-6}\, pc$. For ionized gas of such $EM$ along  every line
of the sight through the halo, as the observation suggests, the free-free 
opacity exceeds unity for $\nu \le 1 \, \rm GHz$. This means  extragalactic sources would not be observable for radio frequencies below $1 \, \rm GHz$. Thus, we can rule out this physical process as being 
responsible for populating the $2s$ state. This inference cannot be changed by clumping the gas along the line of sight
as both $n_{2s}$ and free-free absorption depend upon the square of $n_e$.
 
\subsubsection{Collisions} 
Collisional excitation  can contribute to  
populating the $2s$ state  in  partially ionized gas
at $T \simeq 10^4 \, \rm K$. This gives us:
\be \label{popco}
n_{2s}\simeq {C_{1s,2s}(T) n_e n_{1s}\over A_{2s,1s}} \simeq 4\times  10^{-15} \left ({n_H\over 10 \, \rm cm^{-3}}\right)^2 \, {\rm cm^{-3}}
\ee
For computing the electron density we assume the medium to be collisionally
ionized for $T = 10^4 \, \rm K$. Using $L \simeq N_2/n_{2s}$, we get the 
size of the absorbing column, $L\simeq 50 \, \rm Mpc$ for $n_H \simeq 10 \, \rm cm^{-3}$. Clearly this case results in even more unrealistic values of densities
and the absorbing column. We have checked the whole range of $T$ and $n_{\rm H}$  and could not come up with a single 
case that gives reasonable values such as $n_H \simlt 1 \, \rm cm^{-3}$, as 
observed in the local ISM and in the halo within $L \leq 100 \, \rm kpc$.

\subsubsection{Lyman-series photons} 
If the claimed  H$\alpha$ absorption originates in an extended galactic halo and/or in surrounding IGM gas of the Local group as suggested by (ZZ17), one can 
assume that absorbing gas is exposed to  the extragalactic UV with a photon flux at the Ly-continuum edge $F_\nu^c=10^6~J_{21}$ phot~cm$^{-2}$~s$^{-1}$ \citep[see also 
more recent measurements at $\lambda\approx 2000$ \AA~in][]{frances}. We further  
assume  photon flux at $\lambda\approx 1000 \, \rm  \AA$ is of the same order of magnitude $F_{1000}\sim F_\nu^c$.   We use this value to compute
$\Gamma_{1s,2s} \simeq \sigma_{1s,3p}(T) n_\gamma c (A_{32}/(A_{31}+A_{32})$, where 
$\sigma_{1s,3p}$ is the cross-section of photon absorption from $1s$ to $3p$ level at the line center and $n_\gamma \simeq F_\nu/c$ is the number density of 
photons that cause the transition\footnote{in this case each atom that makes a transition to
the $2s$ state is accompanied by H$\alpha$ emission. However this emission
is isotropic and the flux of these photons along the line of sight is proportional to the solid angle of the observation (roughly square of an arcsecond) which 
is negligible.}.                   
This gives us:
\be 
n_{2s} ={\Gamma_{1s,2s} n_{1s} \over A_{2s,1s}}  \simeq 5\times  10^{-9} \left ({n_H\over 10 \, \rm cm^{-3}}\right) \, {\rm cm^{-3}}
\ee 
For our estimate, we use $T= 5000 \, \rm K$, as expected for the Warm Neutral Medium (WNM) of the ISM and use the absorption cross-section at the line center. 
In this case, we get $L \simeq 400 \, \rm pc$ for $n_H \simeq 1 \, \rm cm^{-3}$. Clearly this case gives a more 
believable picture: a single region of a fraction of the size of the 
galactic halo  with hydrogen  number density and UV flux 
expected in the ISM. 

There are two related issues which need to be discussed before this 
estimate becomes reliable. First, the observed line width, which could be
due to turbulent motion,  is nearly 50~times 
larger than the thermal line width  we assumed here. If we had used the value
of absorption line-center  cross section suitable for the observed line then  the corresponding numbers for $L$ or $n_H$ would be higher by a factor of 50.

\section{Trapped Ly-$\alpha$, metal contamination, DIB, and PAH} \label{sec:dibpah}

There are at least two possible ways to circumvent the constraints in the  foregoing. discussion. The first would be to assume that the $2p$ emission 
originates  from a region  that is optically thick to Ly-$\alpha$, 
the trapping of photons in the region reduces  the effective decay time of  
the $2p$ which also diminishes  the luminosity in the line. The second 
approach would be to posit the observed absorption is caused by a transition
from the ground state of an element other than hydrogen or it arises from 
electronic transitions of  more complex molecules (e.g. DIB carriers or PAH). 

\subsection{Ly-$\alpha$ from optically thick regions}
If the Ly-$\alpha$ emission emerges from a region optically thick to 
this photon, then the coupled problem of solving level populations
along with the evolution of the radiative intensity can be greatly simplified
under the condition of large line center optical depth. In this case  the 
effective $A$-coefficient is replaced by $A_{2p,1s}/(0.5\tau)$, where $\tau$ is 
the optical depth at the line center (e.g. see \cite{2011piim.book.....D}, chapter 19; for a more recent discussion  see  \cite{2016ApJ...820...10D}). In the case,  the occupancy of 
the $2p$ state  increases by $\tau$ and the luminosity in the Ly-$\alpha$ decreases by the same factor. To model this case, we need to ensure 
that the occupancy of $2p$ far exceeds the occupancy of the $2s$ state and 
that the luminosity of the line satisfies the constraints on sky brightness. 
The former can be  achieved by exciting the line with collisions, a case already 
discussed above.  Any transition to states $n >2$ yields a Ly-$\alpha$ photon after a few scattering and therefore the occupation of $2s$ state is suppressed with respect to $2p$ state which gets populated by Ly-$\alpha$ photons 
for which the number of scatterings:   $N_{\rm scat} \simeq \tau$ \footnote{the ratio of the occupancy of $2s$ to $2p$ states scales
as $A_{2s, 1s}/(\tau A_{2p,1s})$, see e.g. \cite{2016ApJ...820...10D}}. To achieve the latter, we need an optical depth $\tau > 10^{10}$, which puts
strong constraints on the HI column. This gives us\footnote{For large optical depth, the level population of the $1s$ and $2p$ states can thermalize or
$n_{2p} = 3 n_{1s} \exp(-h\nu_\alpha/(kT))$.  The equivalent condition is $n_{2p} C_{2p,1s} \simeq A_{2p,1s}/\tau$ and it  is not reached for the range of parameters relevant to the paper.}:
\begin{equation}
n_{2p} \simeq { 0.5 C_{1s,2p}(T) n_e n_{1s} \tau \over A_{2p,1s}}
\end{equation} 
As $n_{2p} \ll n_{1s}$, $n_{1s} \simeq n_H$ in this case. $\tau \simeq \sigma_\alpha(0) n_H L$, where $\sigma_\alpha(0)$ is the cross section for Ly-$\alpha$ at the line center; we assume $T = 10^{4} \, \rm K$ . However, $L \simeq N_2/n_{2p}$ 
to satisfy the observation of H$\alpha$ absorption, which gives:
\begin{eqnarray}
n_{2p} & \simeq & \left ({0.5 C_{1s,2s}(T) n_e n_{H}^2  N_2 \sigma_\alpha(0)  \over A_{2p,1s}} \right )^{1/2} \nonumber \\
& \simeq & 1.3 \times 10^{-10} \left ({n_H \over 10^2} \right)^{3/2} \, \rm cm^{-3} 
\end{eqnarray} 
For $n_H = 10^2 \, \rm cm^{-3}$, the size of the region, $L \simeq 1.7 \, \rm kpc$ and $\tau \simeq 5\times 10^{10}$ and HI column density, $N_{\rm HI} \simeq 5.2 \times 10^{23} \, \rm cm^{-2}$. These numbers are clearly  unrealistic. A change in hydrogen density does not alleviate this problem. Also,  the observed velocity width decreases the line-center cross section by a  factor of 50 which makes the scenario delineated here even less 
plausible.

\subsection{Metal lines}
Given that the width of the  reported  line 
is close to 30~\AA, it is worthwhile to investigate whether this line could arise from metal lines  that lie close to the H$\alpha$ transition.  We consider carbon, nitrogen, and oxygen as they are  the most abundant 
elements in the ISM following hydrogen and helium.   We neglect the lines arising from transitions between two excited states of the atom to avoid 
facing the same issues we discussed in Section \ref{sec:halpha}, e.g., CII 
line at $\lambda \simeq 6578 \, \AA$ which causes a transition between  two excited states.

Two lines of singly ionized nitrogen (NII) ---
$3P_1$--$1D_2$ ($6549.9 \, \AA$) and $3P_2$--$1D_2$ ($6585.3 \, \AA$) --- lie close to H$\alpha$ emission. One distinct advantage in this case is that the absorption is 
caused, unlike H$\alpha$ absorption,  by an element  in its ground state, which means there 
arise no constraints from the decay of the line. 

The abundance of nitrogen in the local ISM is $7 \times 10^{-5}$ (e.g. \cite{2011piim.book.....D}). For simplicity  we assume that all the nitrogen is singly ionized. For $T\simeq 10^{4} \, \rm K$,
the cross-section at the line center for these two transitions, $\sigma_{\rm NII} \simeq 2 \times 10^{-22} \, \rm cm^2$. To achieve the observed optical depth $\simeq 0.01$,  with an average  $n_e \simeq 10^{-2}$,   we require the 
absorbing column to be $L \simeq 30~\rm Mpc$, which is far in the access
of the halo size of the Milky way. Therefore, this can be ruled out as a plausible mechanism to achieve the observed optical depth. 

\subsection{Diffuse interstellar bands} 

If the observed features in absorption around the H$\alpha$ line cannot 
be modelled as either transition of hydrogen or metal then it is conceivable that they arise from more complex molecules. In this section we explore this possibility. 

There are a plethora of  spectral  DIB features. 
These are  presumably caused by carbon chains, PAH and hydrogenated carbons. However, their 
origin remains highly uncertain. Typical DIBs spectral width FWHM$\sim 0.5\hbox{--}3$\AA\, \citep{hobbs08}, is larger than the Doppler 
width for interstellar gas \citep[see Chapter 6 in][]{2010pcim.book.....T}, and it  might mimic the observed wide Doppler width of the 6565\AA\, feature.   

An ongoing ESO survey seeks to detect and characterize DIB for $\lambda=$305--1042~nm with unprecedented spectral resolution
and signal-to-noise ratio, along more than 100 sight-lines \cite{2017arXiv170801429C}. The first results of this survey show a range of spectral features in the 
wavelength band of interest to us \citep[for results on DIB, see also][]{tieliau,2010pcim.book.....T}. It should be pointed out that most of observational data for diffuse interstellar bands (and their templates) in the ISM are obtained from observation of absorption towards nearby stars in the galactic 
plane. Some of these lines of sights show strong H$\alpha$ absorption features that arise  from the stellar atmosphere of the target star, e.g. 
Figure~B1 of \cite{2017arXiv170801429C}. 

\citet{Lan15} stacked SDSS stars, galaxies, and QSOs to detect DIBs in the 
Milky way. Their  composite  absorption spectra,  based on 40000 stellar spectra and obtained after subtracting stellar SEDs, show a discernible absorption feature at 
H$\alpha$ frequency, which they attribute to stellar absorption residual. Even though 95\% of the stars they use have 
temperatures in the range $4500\hbox{--}7000 \, \rm K$, whose spectra are not expected to show prominent H$\alpha$ absorption, a small level of residual absorption   
might remain if the SED of the star is corrected for. We note that if the feature they observe (peak optical depth $\tau \simeq 0.002$ and $\Delta\lambda \simeq 10 \, \rm 
\AA$) is attributed to absorption by ISM, it might  be compatible with the results of (ZZ17). It is because stars compiled by \cite{Lan15} 
lie at typical distance of $2\hbox{--}3 \, \rm kpc$ while the lines of sights analysed by (ZZ17) traverse the galactic halo which could 
be $50 \hbox{--}100 \, \rm kpc$. However,  the errors incurred in subtracting stellar SED might be large enough to remove this weak feature expected from ISM. 

This discussion shows that  there exists an observational selection
bias against the detection of DIB close to H$\alpha$ transition.  
This is supported by observations of emission lines in the Red Rectangle nebula corresponding to DIB absorption lines,
from regions excluding the central star, particularly the strong lines at $\approx 6560$ 
and $6570\,\AA$ \citep{sarre, scar}. Further analysis has shown the presence of two relatively 
strong features at $\lambda=6552.4$\AA\, and $6563.4$\AA\, without 
contamination from H-$\alpha$ line \citep{vanwin}.

This selection effect is avoided in the  observation of absorption from extragalactic 
sources, e.g. SDSS galaxies, for which  intrinsic H$\alpha$ absorption from stellar atmospheres is redshifted, e.g. the work of (ZZ17) or 
for  late type stars without strong H$\alpha$ absorption in their spectra. We 
should also consider laboratory measurements of the spectra of PAH and complex molecules.

\subsection{{PAH lines near $\lambda=6565$\AA} }

From  laboratory measurements of the spectra of complex molecules, we present, as an example, a tentative list of molecular compounds and the corresponding 
lines what are reasonably close (within $30\hbox{--}40$\AA) to the  6565\AA\, and which can in principle mimic  H$\alpha$ absorption.

\begin{enumerate} 
\item {Naphthalene (Np)} cation C$_{10}$H$_8^+$: a feature at $\lambda\simeq 6520$\AA \, ($f=10^{-4}$, $\Delta\nu=100$ cm$^{-1}=42$\AA), 
and a weaker feature at $\lambda\simeq 6600$\AA \, \citep[Table III and Fig. 3a in][]{salama}, vibronic transition from the ground 
state $^2$B$_{3g}$(D$_2$)$\leftarrow$X$^2$A$_u$(D$_0$). 
\item A hydrogenated form H$_n$-HC$_{42}$\,H$_{16}^+$: a feature at 6550\AA\, {($f\approx 0.03$ to $\approx 0.05$)}, \citep[see Fig 5d, e,f in][]{hamm}. 
\item Protonated pyrene 2H-Py$^+$: a relatively strong feature at 6550\AA \, (theoretical oscillator strengths lie 
around $f\sim 0.03-0.05$) \citep[Fig 4 and Table 2 in][]{chin}.  
\item {[FePAH]$^+$ complexes with a band at 6600\AA \, ($f\approx 0.002$), \citep[Table 3 and Fig 2, Fig 3  in][]{lanza}}. 
\end{enumerate}

\subsubsection{Optical depth of DIB/PAH absorbing halo gas}

Let us assume, for  a conservative estimate,  the oscillator strength of the line is  $f_{6565}\simeq 10^{-3}$, the abundance of the carrying compound $\chi_{6565}\simeq 10^{-8}$ \citep[see, e.g.  p. 218 in][]{2010pcim.book.....T} and  the mass of the  compound is $\mu_{6565}$. Then the line-center  optical depth is:
\be \label{pah} 
\tau_{6565}\simeq 8\times 10^{-4}\mu_{6565}^{1/2}T_4^{-1/2} N_{20}
\ee
Here $N_{20} \equiv N_{HI}/(10^{20} \rm cm^{-2})$. 

Similar estimates of typical optical depth for a DIB compound gives,
\be \label{taudib}
\tau_{\rm DIB}={\sqrt{\pi}e^2f\over mc\Delta\nu_{\rm D}}\chi_{\rm DIB}N({\rm H})\simeq 0.08\mu_{\rm DIB}^{1/2}T_4^{-1/2}N_{20}, 
\label{dib}
\ee 
where the
abundance $\chi_{\rm DIB}\simeq 3\times 10^{-10}/f$, with $f$ being the oscillator strength is assumed following \citet{tieliau}. 

From Eqs.~(\ref{pah}) and~(\ref{dib})  we can readily obtain the observed optical depth $\simeq 0.005\hbox{--}0.01$, for an acceptable  range of parameters. 
These estimates are also consistent with constraints from FIR emission \citep[e.g. ][]{2010pcim.book.....T}. The main
reason this constitutes a  plausible  mechanism as compared to metals  is that the absorption cross section is larger for PAH/DIB as 
compared to forbidden lines that lie close the ground state for metals (e.g. NII). 

\section{Summary}  \label{sec:sumcon}

In this paper, we attempt to explain a  recent observation that report the  detection of a broad   
absorption feature ($\Delta \lambda \simeq 30 \, \rm \AA$) at  $\lambda \simeq 6565 \, \rm \AA$   in the galactic halo 
(ZZ17). Even though the feature corresponds to the H$\alpha$ wavelength, we argue  that it could not arise from such a transition. 

 First, the decay of the excited state  through the spontaneous de-excitation of (optically thin) $2p$ and 
$2s$ levels 
gives a sky brightness which is incompatible with UV observations. Second, 
we show that the
observed absorption feature cannot be modelled using $2s$ transition for  parameters expected of the Galactic halo. 
We investigate the possibility of Ly-$\alpha$ photons trapped in an optically thick region. In this case, the sky brightness constraint
can be overcome only for unrealistically  large optical depths.  

We next consider CNO metal lines  close to H$\alpha$ transition. All transitions that connect two excited states can be ruled out (e.g. CII) for the same reason as listed 
above. A doublet of singly ionized nitrogen (NII) has transition frequencies within $20 \, \rm \AA$ of H$\alpha$ transition and the transitions  connect the 
ground state with an excited state, thereby obviating the sky brightness constraint. However, even in this case, we fail to find parameters that are compatible 
with  expected  properties of ISM and Galactic halo. 

Finally, we consider DIB and PAH to explain the observation. There are a number of transitions of such complex molecules in the frequency range of interest. We show that  
known models of DIB and PAH, based on optical and UV absorption of FIR/NIR emission, might  explain the absorption features, even though we are not able to identify the  
molecule responsible for this line. This could partly be owing to the fact that H$\alpha$ absorption in the local stars used to identify these line could 
constitute a selection bias in this case.

Deep observation of a  few bright extragalactic sources (for a range of galactic longitudes) with high spectral resolution (SDSS spectral resolution 
 $\lambda/\Delta\lambda\sim 1500\hbox{--}2500$  could have caused  blending of  lines)   and signal-to-noise ratio might reveal the 
nature of this mysterious absorption feature. 

We thank an anonymous referee for valuable suggestions. YS acknowledges support from RFBR (project code 15-02-08293).

\end{document}